\documentstyle[prl,multicol,aps]{revtex}
\begin{document}

\widetext
\title{High Accuracy Many-Body Calculational Approaches for Excitations in
Molecules}
\author{Jeffrey C. Grossman$^{1,\dag}$, Michael Rohlfing$^{2}$, Lubos Mitas$^{3}$,
Steven G. Louie$^{1}$, and Marvin L. Cohen$^{1}$} \address{${^1}$Department of Physics,
University of California at Berkeley and Materials Sciences Division,
Lawrence Berkeley National Laboratory, Berkeley, CA 94720.\\}
\address{$^2$Institut f\"ur Theoretische Physik II --
Festk\"orperphysik, Universit\"at M\"unster, 
48149 M\"unster, Germany} 
\address{$^{3}$National Center for
Supercomputing Applications, University of Illinois at
Urbana-Champaign, Urbana, Illinois 61801} 
\date{\today} 
\maketitle

\begin{abstract}
Two state-of-the-art computational approaches: quantum Monte Carlo
(QMC), based on accurate total energies, and GW with exciton effects
(GW-BSE), based on perturbation theory are employed to calculate
ionization potentials, electron affinities, and first excited singlet
and triplet energies for the silane and methane molecules.  Results
are in excellent agreement between these dramatically different
approaches and with available experiment.  The optically forbidden
triplet excitation in silane is predicted to lie roughly 1 eV higher
than previously reported.  For methane, the impact of geometry
relaxation is shown to be $\sim$ 2 eV for excited states.  Further, in
the GW-BSE method, we demonstrate that inclusion of off-diagonal
matrix elements in the self-energy operator is crucial for an accurate
picture.
\end{abstract}

\begin{multicols}{2}

Optical excitations play a fundamental role in technological
applications such as dye chemicals, photovoltaics, laser technology,
and catalysts for chemical reactions.  An accurate evaluation of
excited states requires more sophisticated approaches than simply a
straightforward application of the mainstream ground state methods
such as density functional theory and Hartree-Fock.  In addition, many
traditional approaches for calculating excited states can be limited
by a range of factors.  For example, configuration interaction with
single excitations is missing correlation effects and is often
inaccurate, optical transitions modeled as quasiparticle energies
corrected by excitonic effects rely on an approximate description of
the electron-hole interaction, and basis set correlated wave function
approaches are limited by their scaling with the number of correlated
electrons.

Two promising new approaches have been developed to calculate accurate
excitation energies: (1) quantum Monte Carlo (QMC), based on a
stochastic solution to the many-electron Schr\"odinger equation, can
provide chemical accuracy (i.e., 0.05 eV) for the total energy while
at the same time scales as N$^3$ where N is the number of valence
electrons (see, e.g., \cite{Grossman1,Lest97}), and (2) many-body
Green's function perturbation techniques that employ the $GW$
approximation \cite{hedin,hybert} for the electron self-energy
operator to calculate the single-particle excitation spectrum,
followed by a solution to a Bethe-Salpeter equation \cite{strinati}
(BSE) for electron-hole excitations \cite{Rohlfing1}.

In this Letter, we demonstrate excellent agreement between QMC,
GW-BSE, and available experiment for excitation energies of silane
(SiH$_4$) and methane (CH$_4$).  Calculated ionization potentials,
electron affinities, and first excited singlet and triplet energies
also agree very well with highly reliable standard quantum chemistry
approaches; however, the major advantage of our two current approaches
is numerical efficiency.  Both methods require moderate computation
times, and their scaling with system size is to our knowledge superior
to all other quantum-chemistry approaches for computing excited
states.  The excellent agreement between QMC and GW-BSE is encouraging
given the dramatically different methodologies involved (i.e., the
former relies on accurate total energy differences while the latter is
based on perturbation theory).

In our QMC approach \cite{mit91,gro95a,hamm94,pri96,fahy}, 
variational Monte Carlo is employed to find an optimized correlated
many-body trial function. This trial function, $\psi_T(R)$, is a
product of Slater determinants, $D_n$, and a correlation factor
\cite{schm90},
\begin{equation}
\psi_T = \sum_n d_n D_n^\uparrow D_n^\downarrow \exp \left[
\sum_{I,i<j} u(r_{iI},r_{jI},r_{ij}) \right] \; \; ,
\end{equation}
where $I$ corresponds to the ions, $i,j$ to the electrons and
$r_{iI},r_{jI},r_{ij}$ to the relevant separations. Parametrization
and optimization of $u(r_{iI},r_{jI},r_{ij})$, which represents the
electron-electron and electron-electron-ion correlations, is described
in Ref. \cite{mita93}.  In the Slater determinant part, we employ
natural orbitals rather than Hartree-Fock (HF) or density functional
orbitals \cite{gro95a}.  To eliminate most of the remaining
variational bias we use the fixed node diffusion Monte Carlo method, 
which is
based on the property that the operator $\exp(-\tau H)$, where $H$ is
the Hamiltonian, projects out the ground state of any trial function
with the same symmetry and non-zero overlap, subject to the constraint
that the nodes are unchanged \cite{foulkes}. All QMC results
presented here are from the diffusion Monte Carlo approach.

The GW-BSE Green's function perturbation approach consists of three
successive steps.  In the first step a local density approximation
(LDA) \cite{hoh64} calculation is performed to obtain the electronic
ground state.  In the second step the LDA results are used to
construct the electronic self-energy operator $\Sigma$ (also known as
the mass operator).  This is done within the $GW$ approximation
\cite{hedin,hybert} where $G$ is the (LDA) single-particle Green's
function and $W$ is the screened Coulomb interaction based on RPA
dielectric screening.  Solving the equation-of-motion for the
single-particle Green's function yields the quasiparticle states
(i.e., the occupied (hole) states $v$ and the empty (electron) states
$c$) of the molecule. In the third step the electron-hole interaction
$\langle vc | K^{\rm eh} | v'c' \rangle$ between the occupied states
$v$ and the empty states $c$ is evaluated.  The coupled electron-hole
excitations $|S\rangle = \sum_{vc} A^S_{vc} |vc\rangle$ and their
excitation energies $\Omega^S$ then result from the Bethe-Salpeter
equation
\begin{equation}
(\varepsilon^{\rm QP}_c - \varepsilon^{\rm QP}_v ) A^S_{vc} +
  \sum_{v'c'} \langle vc | K^{\rm eh} | v'c' \rangle A^S_{v'c'} =
  \Omega^S A^S_{vc}   \quad .
\end{equation}
For details, we refer to Refs. \cite{Rohlfing1,Rohlfing2}.

We have also calculated excited state energies as total energy
differences using LDA and the generalized gradient approximation (GGA)
with the PW91 functional \cite{per91}. In addition, the small number
of correlated electrons in the present systems permits us to employ
several of the most accurate quantum chemical approaches: coupled
cluster with single, double, and perturbationally triple (CCSD(T))
excitations \cite{Ragh96}, and complete active space self consistent
field (CASSCF) with an active space of 8 electrons and 9 active
orbitals. For these methods, extensive basis set tests showed that
energy differences are converged to better than 0.02 eV with the
cc-pVQZ \cite{Dunning} basis set.  All calculations presented here
other than QMC and GW-BSE have been carried out using the G98 package
\cite{Gaussian} and the cc-pVQZ basis set.

It is important to distinguish between the two different methodologies
employed here.  Using total energy methods, the ionization potential
(IP) and electron affinity (EA) are computed as total energy
differences. In the Green's function, the QP energies are obtained in
a different manner, namely as a solution of a quasiparticle Dyson's
equation \cite{Rohlfing1}.

The single-particle HF and LDA energy spectra for SiH$_4$ are shown in
Fig. \ref{fig_sih4}. While both methods are in qualitative agreement
with one another, there is a large quantitative discrepancy, as
expected: HF usually overestimates HOMO-LUMO gaps and eigenvalue
spacings, whereas LDA typically underestimates these quantities. Note
that since the first excitations are from t$_2$ to a$_1$ orbitals, the
ground and first excited singlet states of SiH$_4$ are of different
symmetry.

It is reassuring that all of the methods (see Table
\ref{tab_qp_energies}) are in good agreement with experiment for the
ionization potential of SiH$_4$.  The LDA, GGA, and HF results are
$\sim$ 0.5 eV lower than experiment, indicating a slight bias towards
the ionized state; however, note the dramatic improvement when
compared to their single-particle highest occupied eigenvalue. The
calculations indicate a small, negative (i.e., nonphysical) electron
affinity ranging from -0.1 to -0.6 eV.

Within the GW-BSE approach, the behavior of the $a_1$ LUMO state
deserves careful discussion (see Table \ref{tab_gw_stuff}).  In the
underlying LDA calculation, the LUMO energy is below the vacuum level
(see Fig. 1) and is thus given by a localized, bound wave function.
Within first-order perturbation treatment of the self-energy operator
$\Sigma$, the orbitals are kept as LDA orbitals and the LUMO energy is
shifted above the vacuum level by 1.7 eV (see second column of
Tab. \ref{tab_gw_stuff}).  Being unbound now, the QP state changes its
wave function from a localized to a more extended orbital.  This is
described by going beyond the diagonal evaluation of $\Sigma$ in the
LDA basis in the $GW$ calculation and taking off-diagonal elements of
$\Sigma$ into account, which are in fact non-zero.  This effect, which
was not accounted for in a previous work \cite{Rohlfing1}, lowers the
LUMO state by as much as 0.8 eV (see third column of
Tab. \ref{tab_gw_stuff}).  The LUMO energy of 0.3 eV is now in good
agreement with the results of the other methods.  Note that the LUMO
state (being unbound) converges slowly with respect to the basis, due
to its delocalized nature.  The low-energy electron-hole excited
states, on the other hand, that are the focus of this paper, are
spatially localized since the electron is attracted to the hole; these
excited states are much easier to converge.

Our results for SiH$_4$ excitation energies for the lowest-energy
triplet and singlet excited states are given in Table
\ref{tab_excite_energies}. Calculation of spin-singlet excitations by
total energy differences poses a serious challenge to traditional
approaches. For example, the first excited singlet state is described
by a 2 determinant wavefunction. Accurate correlation of this state
was therefore only possible using the CASSCF method.  The triplet
state is significantly easier to converge within traditional
approaches owing to the fact that it is the lowest energy state of the
given spin multiplicity. The QMC approach does not encounter the same
difficulties as the traditional approaches for the singlet state since
the spin configuration is given by specification of the slater
determinant -- singlet and triplet states are equally as simple to
simulate, provided each wavefunction is orthogonal to the ground
state. In the case of the excited singlet state, the DMC energy was
found to be rather insensitive (i.e., $<$0.1 eV) to the number of
determinants in the trial function, indicating that the nodes of a
single determinant closely resemble those of the 2-determinant
wavefunction which has the correct spatial symmetry.

The SiH$_4$ spin singlet results are in good agreement with experiment
\cite{sih4_exp} for the three theoretical methods listed in Table
\ref{tab_excite_energies}.  The optically forbidden triplet excitation
is difficult to obtain from experiments.  Using electron collision
spectroscopy, slight spectral structures were found at around 7-8 eV
\cite{trip_exp}.  We believe that the agreement among the theoretical
methods presented here strongly suggests that the triplet excitation
lies roughly 1 eV higher than previously reported.

The GW-BSE results of Table \ref{tab_excite_energies} differ from the
ones of Ref. \cite{Rohlfing1} (see also Table \ref{tab_gw_stuff}). The
reason is again due to the off-diagonal matrix elements of the
self-energy operator when expanded in the LDA orbitals, which had been
neglected in Ref. \cite{Rohlfing1}. As discussed above, the full
self-energy operator allows the unoccupied states to become more
delocalized. Therefore the overlap between the hole and electron
states, as well as the electron-hole interaction, is significantly
reduced.  Concomitantly, the excitation energies (column 3 in
Tab. \ref{tab_gw_stuff}) are not reduced compared to column 2 (as the
lower LUMO energy suggests), but are in fact increased.  It is
interesting to note that the triplet excitation energy is more
affected than the singlet, which reduces the singlet-triplet splitting
from 1.1 eV to 0.7 eV within GW-BSE.

Methane and silane are isoelectronic and similar in their valence
shell structure; however, the stronger potential of carbon compared to
silicon causes wavefunctions and densities to be less smooth which in
turn can lead to difficulties and inaccuracies within ab initio
calculations.  We have carried out the same calculations described
above for CH$_4$. Our results, shown in Table \ref{tab_ch4}, again
demonstrate strong agreement between QMC and GW-BSE.

An important issue in the discussion of molecular spectra is the
interaction between electronic excitations and the molecular geometry.
Despite the high dimensionality of a five-atom molecule, we are able
to discuss the most important aspects of geometric relaxation in the
excited states and their consequences for the total energy.  Table
\ref{tab_ch4} illustrates the impact of relaxation from the ground
state T$_d$ symmetry for each excited state molecule.  Note that the
optimization has been restricted to C$_{2v}$ symmetry.  Test
calculations show that further lowering the symmetry results in a
total energy reduction no larger than $\sim$ 0.1 eV.

For all three excited states shown here, geometry relaxation yields an
enormous lowering of the total energy by as much as 1.9 eV.  A minor
contribution (about 0.2-0.4 eV) is related to the increase of the C-H
distance in the excited states.  The main contribution results from
breaking the $T_d$ symmetry of the ground state.  This effect can be
understood from the molecular orbital scheme (see
Fig. \ref{fig_sih4}).  Within the ground state, all three degenerate
HOMO levels are fully occupied, i.e. the ground state is
non-degenerate.  In the excited states, one of the three HOMO states
becomes half-filled, i.e. the excited states are three-fold
degenerate.  The symmetry-breaking from $T_d$ to $C_{2v}$ lifts the
degeneracy of the three HOMO levels.  In the single-particle spectrum
of Fig. \ref{fig_sih4} two HOMO levels are shifted down and one is
shifted up, forming the new HOMO level of the molecule in the reduced
symmetry.  In the optimum geometry of the excited states this
level-splitting is about 4.2 eV, although the center-of-mass of the
three levels remains roughly the same. The new HOMO level, which is
half-filled in all three excited states, is non-degenerate, so the
excited states are non-degenerate as well.

A closer analysis of the excited-state geometry shows that the
increase of the C-H bond length is relatively small (no more than 0.07
\AA).  The main effect results indeed from the Jahn-Teller distortion
of the molecule, which changes the bond angles from 109.47 degree
(tetrahedral angle) to 94 degree for the small and 118 degree for the
large bond angles.  This distortion increases the total energy of the
ground state by 1.5 eV.  In the excited states, however, this increase
of the total energy is overcompensated by the splitting of the three
HOMO levels described above, i.e. by the upwards shift of the new HOMO
level and a correspondingly large reduction of the excitation
energies.  In the experimental spectrum for CH$_4$ \cite{herzberg}, a
weak shoulder is observed at about 10 eV, corresponding to a vertical
excitation.  Compared with the onset at 8.52 eV, this indicates a
relaxation gain of about 1.5 eV, supporting our calculated results.

Similar effects on the excitation energy are also found for SiH$_4$.
The gain in total energy is 0.9 eV, i.e. slightly smaller than in
CH$_4$.  In the measured absorption spectrum this corresponds to the
difference between the energy of maximum absorption and the low-energy
onset.  This difference amounts to 0.6 eV in the absorption
measurements of Ref. \onlinecite{sih4_exp}.  Our minimum-energy
transition for SiH$_4$ is 8.3 eV (in GW-BSE), in excellent agreement
with the measured onset of the spectrum at 8.2 eV.

JCG acknowledges useful discussions with G.M. Rignanese and E. Chang.
Support was provided by National Science Foundation Grant No.
DMR-9520554 and by the Director, Office of Energy Research, Office of
Basic Energy Sciences, Materials Sciences Division of the
U. S. Department of Energy under Contract No. DE-AC03-76SF00098.  MR
acknowledges financial support by the Deutsche Forschungsgemeinschaft
(Bonn, Germany) under grant No. Ro 1318/2-1.  Computational resources
have been provided by NCSA and by NERSC.

$\dag$ Present Address: Lawrence Livermore National Laboratory,
7000 East Avenue, L-415, Livermore, CA 94550.


\narrowtext
\begin{figure}[ht]
\caption{Single-particle Hartree-Fock and local density approximation
eigenvalue spectra (eV) for the SiH$_4$ molecule.
Note that the CH$_4$ molecule has the same orbital structure.
}
\label{fig_sih4}
\end{figure}

\narrowtext
\begin{table}[ht]
\begin{center}
\caption{Ionization potential (IP) and quasiparticle (QP) gap (eV) for
SiH$_4$, computed as E(N-1)-E(N) and E(N+1) + E(N-1) - 2E(N),
respectively, for all methods except GW.
}
\bigskip
\begin{tabular}{l l l}
  & IP
  & QP Gap \\
\hline
HF       & 12.3  & 12.6        \\
LDA      & 12.1  & 12.4        \\
BPW91    & 12.1  & 12.5        \\
CCSD(T)  & 12.7  & 13.3        \\
GW       & 12.7  & 13.0        \\
QMC      & 12.6(1)  & 12.8(1)  \\
EXP$^a$  & 12.6  & ---         \\
\hline
\end{tabular}
$^a$Ref. \onlinecite{sih4_exp}.
\label{tab_qp_energies}
\end{center}
\end{table}

\begin{table}
\caption{
Quasiparticle levels and lowest electron-hole excitation energies
(eV), calculated for SiH$_4$ within LDA and within two evaluations of
the GW-BSE approach.  Unlike the second column, the GW-BSE results of
the third column includes off-diagonal matrix elements of the
self-energy operator and allows for changes in the one-particle wave
functions.  }
\vspace*{2ex}
\label{tab4}
\begin{tabular}{lrrr}
             &  LDA  &  GW-BSE (diagonal $\Sigma$) &  GW-BSE (full $\Sigma$)\\
\hline
HOMO energy  & -8.4  &   -12.7        &      -12.7\\
LUMO energy  & -0.6  &     1.1        &        0.3\\
$E_S$        &  7.8  &     8.8        &        9.2\\
$E_T$        &  7.8  &     7.7        &        8.5\\
\end{tabular}
\label{tab_gw_stuff}
\end{table}

\narrowtext
\begin{table}[ht]
\begin{center}
\caption{Neutral excitation energies (eV) for the first excited singlet and
triplet states of SiH$_4$ computed as E(excited state) - E(ground state)
for all methods except GW-BSE. $\Delta$ corresponds to the
singlet-triplet splitting.
The experimental result \protect\cite{sih4_exp} denotes the energy of maximum 
absorption for the state studied here, which corresponds to vertical 
excitation.}
\bigskip
\begin{tabular}{l l l l}
  & singlet
  & triplet
  & $\Delta$ \\
\hline
HF       & ---    & 8.4    & ---  \\
LDA      & ---    & 8.1    & ---  \\
BPW91    & ---    & 8.2    & ---  \\
CCSD(T)  & ---    & 8.7    & ---  \\
CASSCF   & 9.1    & 8.7    & 0.4  \\
GW-BSE       & 9.2    & 8.5    & 0.7  \\
QMC      & 9.1(1) & 8.7(1) & 0.4(1) \\
EXP      & 8.8    & ---    & ---  \\
\hline
\end{tabular}
\label{tab_excite_energies}
\end{center}
\end{table}

\narrowtext
\begin{table}[ht]
\begin{center}
\caption{Ionization potentials (IP) and energies of the lowest 
singlet (E$_S$)
and triplet (E$_T$) excited states
for CH$_4$ in both the ground state (T$_d$) and relaxed
(C$_{2v}$) symmetries. All energies are in eV and are given
with respect to the ground state.
The experimental result \protect\cite{herzberg} denotes the onset of the 
spectrum, which corresponds to the minimum energy transition including
structural relaxation.
}
\bigskip
\begin{tabular}{l l l l}
  & IP
  & E$_T$ 
  & E$_S$ \\
\hline
GW-BSE (T$_d$)    & 14.3       & 10.1    & 10.5    \\   
QMC (T$_d$)       & 14.3(1)    & 10.1(1) & 10.4(1) \\
GW-BSE (C$_{2v}$) & 12.5       & 8.2     & 8.6     \\
QMC (C$_{2v}$)    & 12.7(1)    & 8.4(1)  & 8.7(1)    \\
EXP               & 12.99$^a$  & ---     & 8.52    \\
\hline
\end{tabular}
$^a$ Ref. \onlinecite{herzberg}
\label{tab_ch4}
\end{center}
\end{table}

\end{multicols}

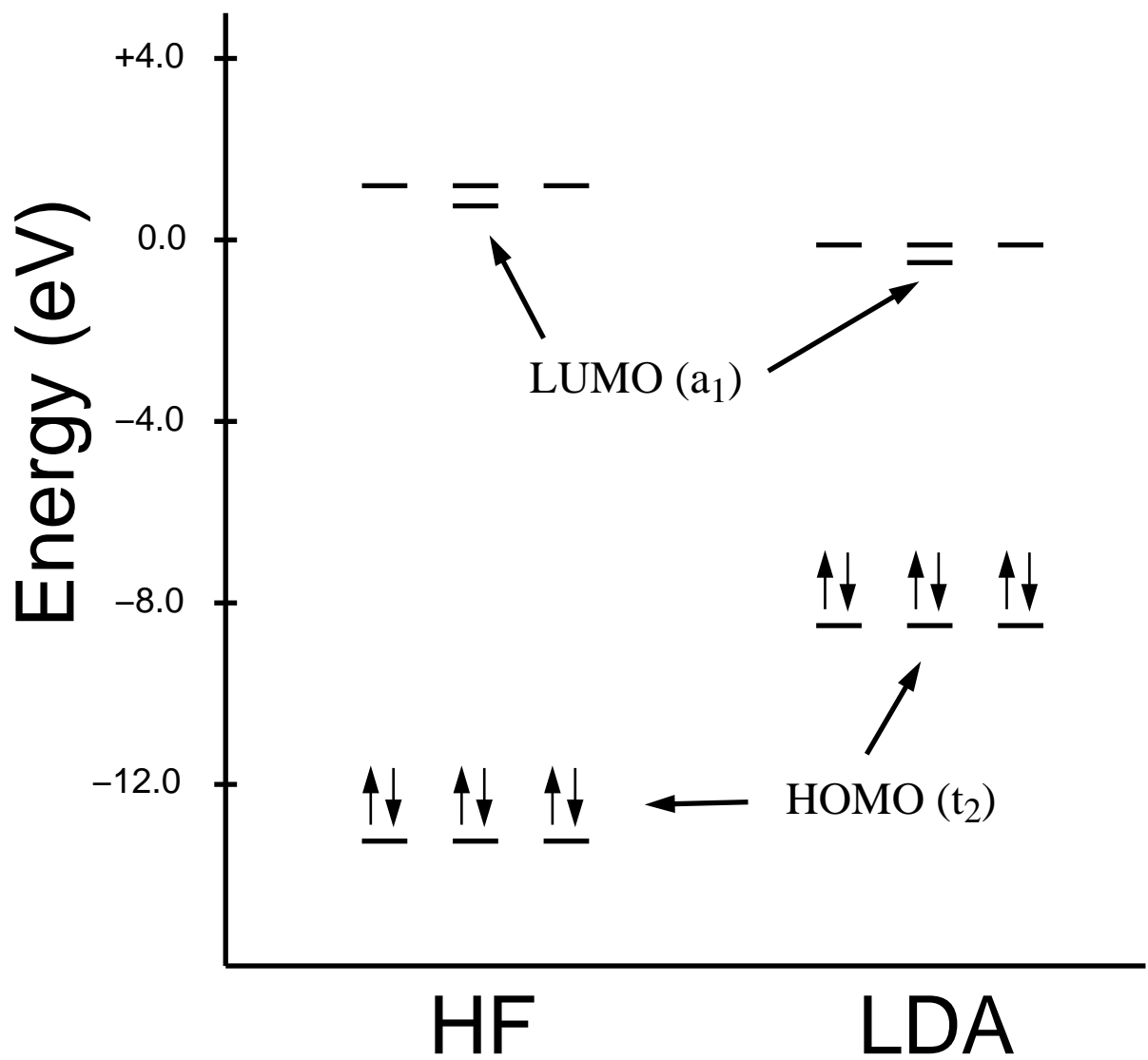
\end{document}